\documentclass{elsart}
\usepackage{psfig}
\usepackage{natbib}
\begin{document}

\runauthor{Cicero, Caesar and Vergil}

% -------------------------------------------------------------------------

\begin{frontmatter}

\title{The response of the X-ray luminosity to the interaction strength in Seyfert 1 \& 2 galaxies}

\author[PSU]{F. Pfefferkorn \& T. Boller}
\address[PSU]{Max-Planck-Institut f\"ur extraterrestrische Physik, Giessenbachstrasse, 85748 Garching, 
Germany email: pfefferk@mpe.mpg.de \& bol@mpe.mpg.de}

\begin{abstract}

We have correlated a sample of 99 Seyfert 1 \& 98 Seyfert 2 (Rafanelli et al., 1995) interacting
galaxies with the survey and pointed ROSAT data and have detected 
91 Seyferts 1 \& 47 Seyfert 2s in the X-ray band. 
We have performed spectral and timing analysis of the X-ray detected sources, 
and we have examined the relation of the 
X-ray luminosity $L_X$ with the interaction strength $Q$.

\end{abstract}


\end{frontmatter}

% -------------------------------------------------------------------------

\section{Introduction}

Interacting galaxies are considered to have a wide importance on triggering of starburst and AGN activity.
Therefore, we have investigated two subsamples of spectroscopically-selected, 
X-ray detected interacting Seyfert 1s, 
including NLS1, and Seyfert 2s galaxies (Rafanelli et al.,1995) with 
$ z<0.11, \; m \le 15.5 \; and \; \delta \ge -23^\circ $.\\
The pairs have been selected on POSS plates using the following criteria 
given by Rafanelli et al. (1995): separation between the components
$S < 3D_p$, where $D_p$ is the apparent major axis of the\
Seyfert galaxy, and a magnitude-difference between Seyfert and the companion\
$\Delta m = m_{comp} - m_{seyfert} < 3$ .

The identification is based on X-ray contour lines overlaid on optical images taken from DSS.
In Fig. 1 we show as an example the overlay from the NLS1 galaxy Mkn 896. 
Fig. 2 shows the spectrum of this object.
The deviation from the fit at energy 0.7 keV in Fig. 2 indicates the presence of a warm absorber.
The pointing and survey lightcurves in Fig. 3 show the variability of this source on short time scales.

\begin{figure}[htb]
\centerline{\psfig{figure=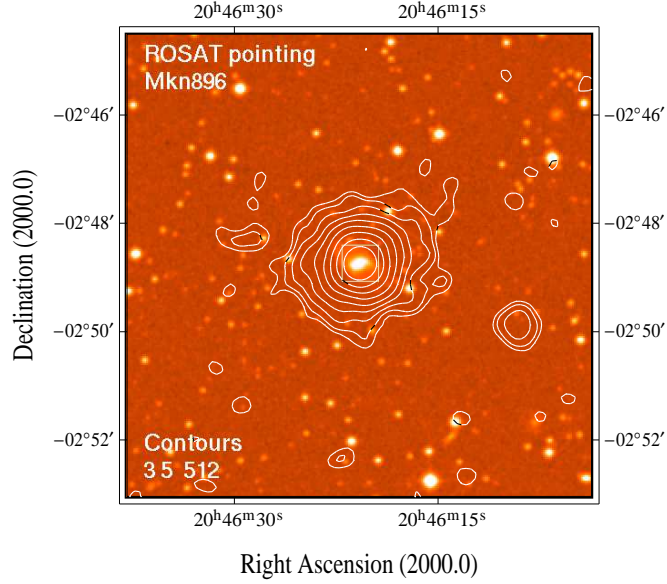,height=3.1truein,width=3.1truein,angle=0}}
\caption{The NLS1 galaxy Mkn 896 with a companion galaxy. The X-ray 
contours are overlaid on the optical image.}
\end{figure}

\begin{figure}[htb]
\centerline{\psfig{figure=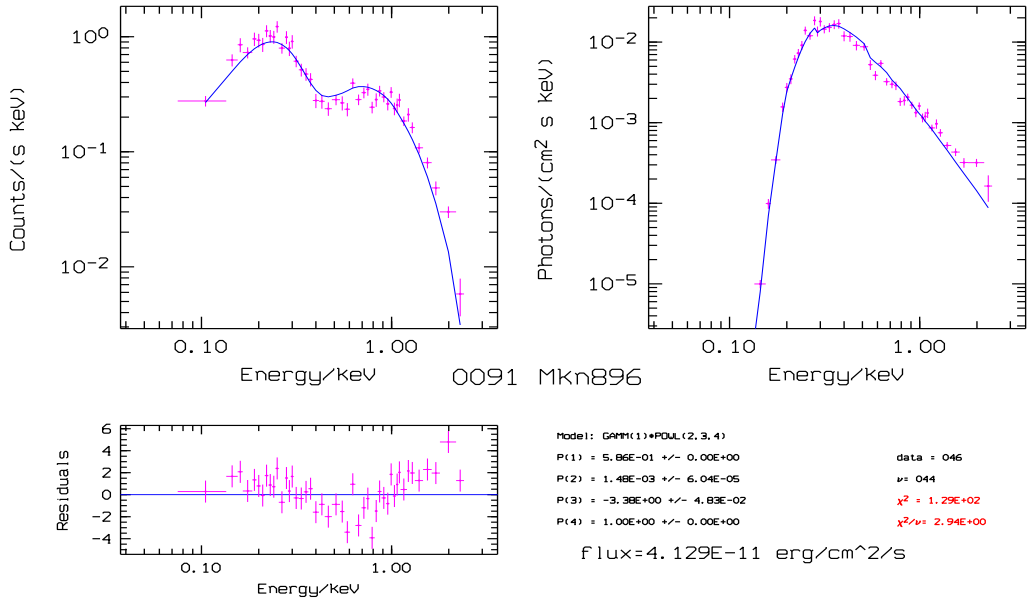,height=3.3truein,width=5.6truein,bbllx=25mm,bblly=10mm,bburx=200mm,bbury=270mm,angle=-90,clip=}}
\caption{Simple power-law fit results to the NLS1 Mrk 896.}
\end{figure}

\begin{center}
\begin{figure}[htb]
\hbox{
\psfig{figure=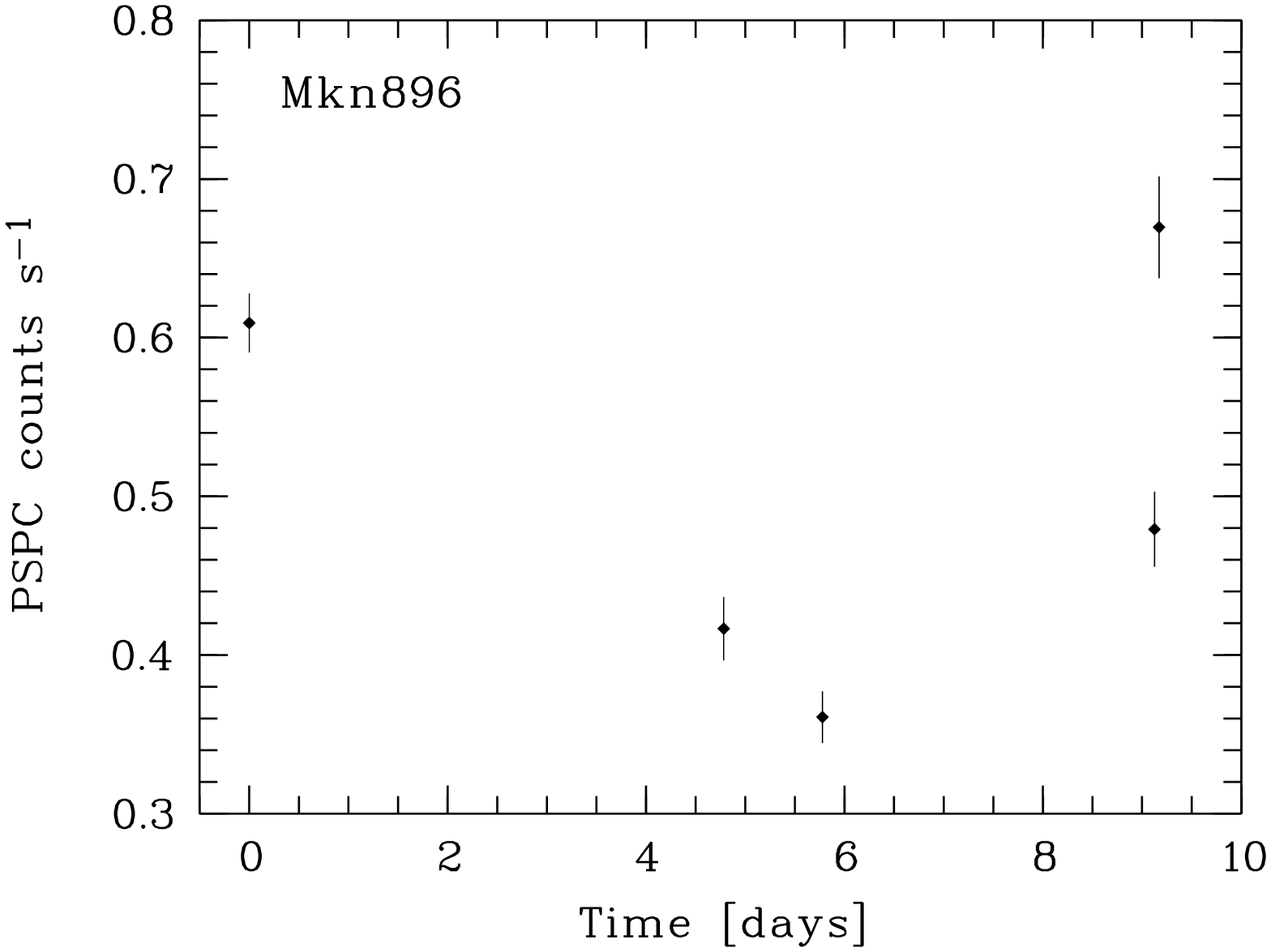,height=2.5truein,width=2.8truein,bbllx=10mm,bblly=20mm,bburx=190mm,bbury=240mm,angle=-90,clip=}
\psfig{figure=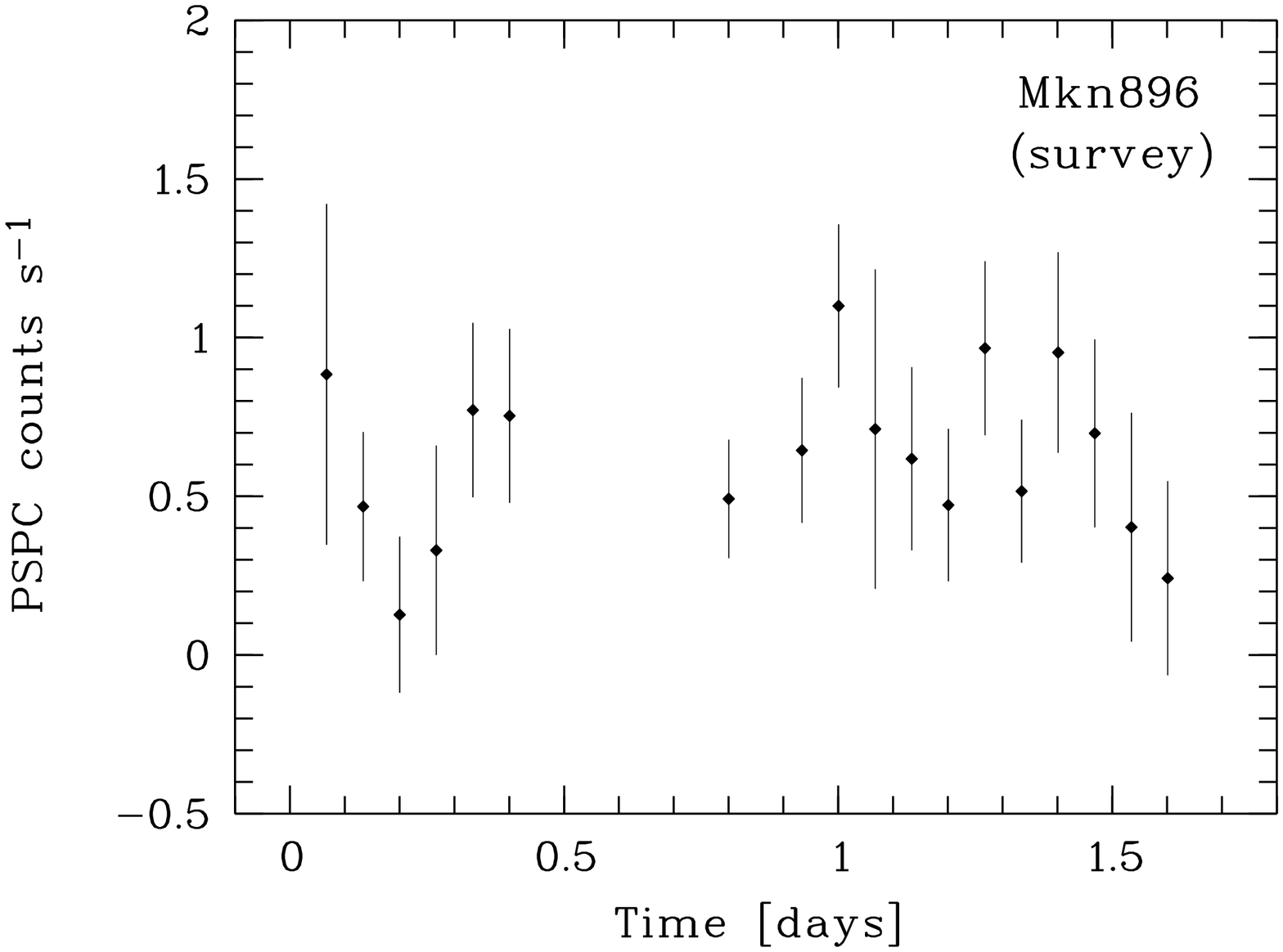,height=2.5truein,width=2.6truein,bbllx=10mm,bblly=35mm,bburx=190mm,bbury=240mm,angle=-90,clip=}
}
\caption{The lightcurves from the pointing and survey observation shows the short-time variability of Mkn896.}
\end{figure}
\end{center}

\section{First results on the interaction strength $Q$}

To derive the interaction strength $Q$, we concentrate on the tidal force per unit mass produced
by a companion on a primary galaxy, which is proportional to 
$M_c \cdot R^{-3}$. $M_c$ is the
mass of the companion and $R$ is its distance from the center of 
the primary galaxy. In most
cases, $M_c$ and the absolute value of $R$ are unknown. 
Instead, these parameters are related to the dimensions
of the pair. Rubin et al. (1982) describe the dependence of the mass $M$ of a galaxy on the size
of its major axis $d$ as $M \propto d^{\gamma}$ and we use $\gamma = 1.5$ (Dahari, 1984). If we
utilize the apparent diameter of the primary galaxy $D_p$ as a scaling 
factor, we obtain: 
\vspace{-0.3cm}
\[ M_c \propto (D_c/D_p)^{1.5} \qquad and \qquad R \propto S/D_p \]
\vspace{-0.5cm}
Using these relations, we get as the dimensionless gravitational interaction strength $Q$:
\vspace{-0.3cm}
\[ \frac{M_c}{R^3} \propto \frac{(D_c \cdot D_p)^{1.5}}{S^3} \equiv Q \]
\vspace{-0.4cm}
This parameter is obviously large for close and relatively large companions.

\begin{figure}[htb]
\centerline{\psfig{figure=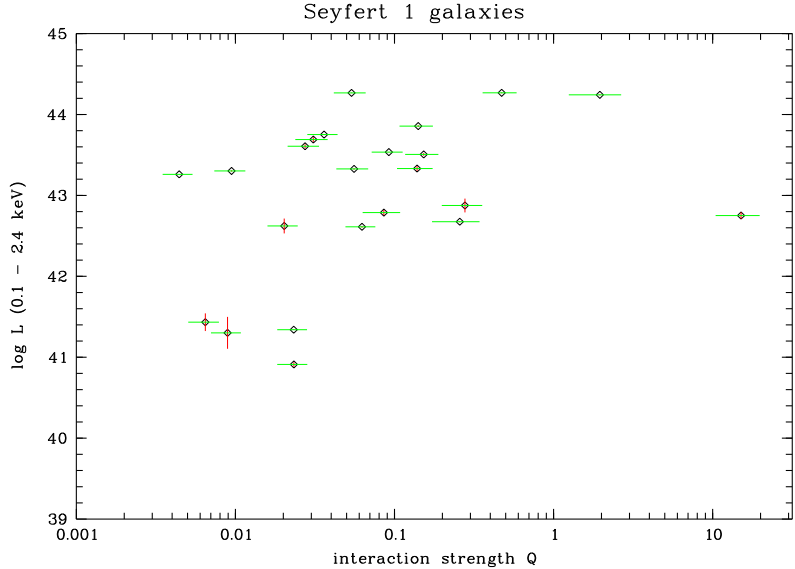,height=3.2truein,width=4.7truein,angle=-90}}
\caption{Soft X-ray luminosity L vs. dimensionless gravitational 
interaction strength Q for  the
Seyfert 1 galaxies. The plot suggests an increase in luminosity for increasing values of the
interaction strength. The luminosity has been calculated from the absorption corrected flux,
integrated over the energy range $0.1 - 2.4 \;keV$ and using the Hubble constant
$H_o=75 \;\frac{km}{\;s\;Mpc}$.}
\end{figure}

Fig.4 shows interaction strength $Q$ vs. the X-ray luminosity for Seyfert 1 galaxies.
Noticeable is a faint tendency for the  luminosity increase for increasing values of interaction
strength.
A possible explanation for the low luminosity sources are the presence of a high hydrogen
column density in the nucleus and the emission from the circumnuclear starbursts.\\
The galaxy interactions or mergers are able to trigger a burst of star formation, which can
be a reason for the rise in luminosity. Several studies about physical 
mechanisms which could
trigger a burst of strong star formation have been investigated by Barnes \& Hernquist (1991),
Jog \& Das (1992) and Jog \& Solomon (1992).
Also many observations suggest that galactic encounters enhance star formation rates, as
demonstrated by studies of the optical (e.g., Kennicut et al. 1987; Bushouse 1987), infrared
(e.g., Bernl\"ohr 1993; Telesco et al. 1993) and radio (e.g., Hummel 1981; Smith \& Kassim 1993)
emission from interacting systems.\\
For each source detected in the All-Sky Survey with more than 100 counts in an exposure time of 400 sec
and for all sources within pointed observations, we performed a spectral analysis and created a lightcurve. Studies of variability were possible,
if more than one pointed observation was available or if both a 
detection in the survey and in a pointed observation existed.
We are extending our studies on starburst galaxies to examine the influence of
interaction with companions on the X-ray luminosity. A paper with data concerning X-ray properties,
luminosity, flux, count rate variation and lightcurves for Seyfert 1s \& 2s is in preparation.

% -------------------------------------------------------------------------

% -------------------------------------------------------------------------

\end{document}